\documentclass{emulateapj}
\usepackage{epsfig}

\slugcomment{Accepted for publication in the Astrophysical Journal}

\shorttitle{Chandra Observation of A13}
\shortauthors{JUETT ET AL.}

\begin{document}

\title{A {\em Chandra} Observation of Abell 13: Investigating the
Origin of the Radio Relic}

\author{Adrienne~M.~Juett\altaffilmark{1},
Craig~L.~Sarazin\altaffilmark{1}, Tracy~E.~Clarke\altaffilmark{2},
Heinz~Andernach\altaffilmark{3}, Matthias~Ehle\altaffilmark{4},
Yutaka~Fujita\altaffilmark{5}, Joshua~C.~Kempner\altaffilmark{6},
Alan~L.~Roy\altaffilmark{7}, Lawrence~Rudnick\altaffilmark{8}, and
O.~Bruce~Slee\altaffilmark{9}}

\altaffiltext{1}{Department of Astronomy, University of Virginia,
P.O. Box 400325, Charlottesville, VA 22904-4325, USA;
ajuett@virginia.edu, sarazin@virginia.edu}
\altaffiltext{2}{Naval Research Laboratory, 4555 Overlook Ave. SW,
Code 7213, Washington, DC 20375, USA; Interferometrics Inc., 13454
Sunrise Valley Drive, Suite 240, Herndon, VA 20171, USA}
\altaffiltext{3}{Departamento de Astronom\'{\i}a, Universidad de
Guanajuato, AP 144, Guanajuato CP 36000, Mexico}
\altaffiltext{4}{XMM-Newton Science Operations Centre, European Space
Agency, Villafranca del Castillo, P.O. Box 50727, 28080 Madrid, Spain}
\altaffiltext{5}{Department of Earth and Space Science, Graduate
School of Science, Osaka University, Toyonaka, Osaka 560-0043, Japan}
\altaffiltext{6}{Department of Physics and Astronomy, Bowdoin College,
8800 College Station, Brunswick, ME 04011, USA}
\altaffiltext{7}{Max-Planck-Institut f\"{u}r Radioastronomie, Auf dem
H\"{u}gel 69, D-53121 Bonn, Germany} 
\altaffiltext{8}{University of Minnesota, 116 Church St SE,
Minneapolis, MN 55455, USA}
\altaffiltext{9}{Australia Telescope National Facility, P.O. Box 76,
Epping, NSW 2121, Australia}

\begin{abstract}
We present results from the {\em Chandra} X-ray observation of Abell
13, a galaxy cluster that contains an unusual noncentral radio source,
also known as a radio relic.  This is the first pointed X-ray
observation of Abell 13, providing a more sensitive study of the
properties of the X-ray gas.  The X-ray emission from Abell 13 is
extended to the northwest of the X-ray peak and shows substructure
indicative of a recent merger event.  The cluster X-ray emission is
centered on the bright galaxy H of \citet{srm+01}.  We find no
evidence for a cooling flow in the cluster.  A knot of excess X-ray
emission is coincident with the other bright elliptical galaxy F.
This knot of emission has properties similar to the enhanced emission
associated with the large galaxies in the Coma cluster.

With these {\em Chandra} data we are able to compare the properties of
the hot X-ray gas with those of the radio relic from VLA data, to
study the interaction of the X-ray gas with the radio emitting
electrons.  Our results suggest that the radio relic is associated
with cooler gas in the cluster.  We suggest two explanations for the
coincidence of the cooler gas and radio source.  First, the gas may
have been uplifted by the radio relic from the cluster core.
Alternatively, the relic and cool gas may have been displaced from the
central galaxy during the cluster merger event.   
\end{abstract}


\section{Introduction}
The high spatial resolution of the {\em Chandra X-ray Observatory}
allows us to study the intracluster medium (ICM) in clusters of
galaxies in unprecedented detail.  One interesting result is the
interplay between the X-ray emitting gas and non-thermal radio
emission.  Observations of several galaxy clusters have found X-ray
cavities in the ICM coincident with bright radio emission \citep[e.g.,
Perseus (A\,426);][]{bvf+93,fse+00}.  The central radio sources
produce the bubbles of non-thermal plasma responsible for the radio
emission.  These bubbles then rise, possibly by buoyancy, displacing
the X-ray emitting gas and causing cavities in the latter.

A number of galaxy clusters contain large-scale diffuse radio sources
not associated with active galactic nuclei.  These sources are
classified into two groups: radio halos which are centrally located,
and radio relics.  Radio relics are extended diffuse radio sources
often located at the periphery of clusters of galaxies.  \citet{gf04}
present a classification scheme for radio relics based on their
observational properties \citep[but see also][]{kbc+04}.  Radio relics
typically have steep radio spectra ($\alpha > 1$ assuming $S_{\nu}
\propto \nu^{-\alpha}$) and are linearly polarized \citep{gf04}.  The
X-ray surface brightness around most radio relics is low, making it
difficult to study the relationship between the X-ray gas and radio
emission.

\begin{deluxetable*}{lccc}
\tablewidth{0pt}
\setlength{\tabcolsep}{0.2in}
\tablecaption{Properties of Detected Point Sources in Abell 13 Field}
\tablehead{\colhead{Name} & \colhead{R.A.\tablenotemark{a}} & 
\colhead{Dec.\tablenotemark{a}} & \colhead{Count Rate ($10^{-3}$~s$^{-1}$)} }
\startdata
CXOU J001316.6$-$193201   &   00:13:16.67 &  $-$19:32:01.2  &  6.3$\pm$0.4 \\
CXOU J001319.0$-$193018   &   00:13:19.09 &  $-$19:30:18.5  &	0.43$\pm$0.12 \\
CXOU J001320.0$-$192748   &   00:13:20.06 &  $-$19:27:48.6  &  2.1$\pm$0.2 \\
CXOU J001321.9$-$192750   &   00:13:21.98 &  $-$19:27:50.7  &	2.4$\pm$0.2 \\
CXOU J001326.0$-$193111   &   00:13:26.02 &  $-$19:31:11.4  &	3.9$\pm$0.3 \\
CXOU J001335.7$-$192805   &   00:13:35.74 &  $-$19:28:05.5  &	0.61$\pm$0.14 \\
CXOU J001339.9$-$192935   &   00:13:39.94 &  $-$19:29:35.8  &	4.2$\pm$0.3 \\
CXOU J001341.1$-$192809   &   00:13:41.15 &  $-$19:28:09.5  &	1.8$\pm$0.2 \\
CXOU J001342.5$-$193209   &   00:13:42.58 &  $-$19:32:09.0  &  0.54$\pm$0.13 \\
CXOU J001344.6$-$192914   &   00:13:44.60 &  $-$19:29:14.1  &	0.38$\pm$0.11 \\
CXOU J001344.7$-$193051   &   00:13:44.70 &  $-$19:30:51.0  &	4.6$\pm$0.3 
\enddata
\tablenotetext{a}{Units of right ascension are hours, minutes, and
seconds, and units of declination are degrees, arcminutes, and
arcseconds.  The 90\% uncertainty in absolute positional accuracy of
{\em Chandra} is 0\farcs6.}
\label{tab:srcprop}
\end{deluxetable*}

Many radio relics, including some of the best known and studied
\citep[e.g., the Coma cluster relic 1253$+$275 and Abell
3667;][]{gfs91,rwh+97}, are located far from the cluster center, on
the order of Mpcs.  However, there exists a subclass of radio relics
which are located significantly closer to the cluster center, $\sim$50
to 350 kpc.  \citet{gf04} classify these as relic sources near the
first ranked galaxy.  These radio sources share many of the same
properties of classical radio relics including steep spectra,
filamentary structure, and polarization \citep{srm+01}.  Abell 133 is
one of the class of radio relics near the first ranked galaxy.  A {\em
Chandra} observation revealed a tongue-like feature in the X-ray
emission comprised of cool gas ($kT \approx 1.3$~keV) which extended
from the cD galaxy to the radio relic \citep{fsk+02}.  A number of
possible origins for the tongue were explored including the suggestion
that the cool gas in the tongue was uplifted from the cluster core by
the buoyantly rising radio relic.  A followup observation with {\em
XMM-Newton} is consistent with the uplifted bubble scenario
\citep{fsr+04}.  However the detection of a cold front southeast of
the cluster core and a weak shock in the core suggests that an unequal
mass merger is responsible for the tongue feature.  In either model,
the radio plasma appears to have originated from the active galactic
nucleus (AGN) in the central cD, rather than as a direct result of
merger shock acceleration.

In this paper, we present X-ray data obtained with the {\em Chandra
X-ray Observatory} for the cluster Abell 13, another unusual system
with a radio relic near the first ranked galaxy.  Abell 13 is an X-ray
luminous ($8\times10^{43}$~erg~s$^{-1}$ in the 0.5--10~keV range)
cluster at $z=0.0943$ \citep{mkd+96}.  The relic has an extremely
steep ($\alpha \approx 4$) radio spectrum, a linear polarization of
$\approx$12\% at 1425~MHz, and spans up to 2\arcmin\/ \citep{srm+01}.
No identification of a host galaxy associated with the relic was
possible, although the two brightest cluster members are candidates.
Interestingly, these two galaxies are separated by 2621~km~s$^{-1}$ in
recessional velocity \citep{qr95}.  Similarly, from redshift data of
the ESO Nearby Abell Clusters Survey (ENACS), \citet{fgg+96} concluded
that the radial velocity distribution of Abell 13 may also be
interpreted as a bimodal one, with peaks at $z=0.0919$ (16 redshifts,
vel.disp. $361^{+53}_{-35}$~km~s$^{-1}$) and $0.0972$ (21 redshifts
vel.disp. $515^{+104}_{-81}$~km~s$^{-1}$), or $<cz> = 27570$ and
29160~km~s$^{-1}$.  We assume $H_0=70$~km~s$^{-1}$~Mpc$^{-1}$,
$\Omega_M=0.3$, and $\Omega_\Lambda=0.7$ unless otherwise mentioned.
At the distance of Abell 13, 1\arcsec\/ corresponds to 1.75 kpc.

\section{Spatial Analysis}

We observed Abell 13 on 2004 August 25 for 55~ks with {\em Chandra}
using the Advanced CCD Imaging Spectrometer \citep[ACIS;][]{gbf+03}.
The observation was performed so that both the cluster center and the
radio relic were located on the back-illuminated S3 CCD.  The ``level
1'' event file was processed using the CIAO v3.2 data analysis
package\footnote{http://asc.harvard.edu/ciao/} following the standard
reduction procedure.  We removed the afterglow correction and created
a new observation specific bad pixel file.  The level 1 event file was
reprocessed to include the new bad pixel file and utilize the improved
background event detection in VFAINT mode.  We then filtered the data
to remove hot pixels, bad columns, and events with grades 1, 3, 5, and
7.  Data from the back-illuminated S1 CCD were used to determine time
intervals with high background using the tool {\tt lc\_clean}.  We
filtered the event file on good times, yielding a final exposure time
of 50.7~ks.  Background data were extracted from the blank sky
observations compiled by M. Markevitch\footnote{See
http://cxc.harvard.edu/cal/Acis/Cal\_prods/bkgrnd/acisbg/}.  Our
analysis is restricted to data from the S3 CCD.

\begin{figure*}
\centerline{\epsfig{file=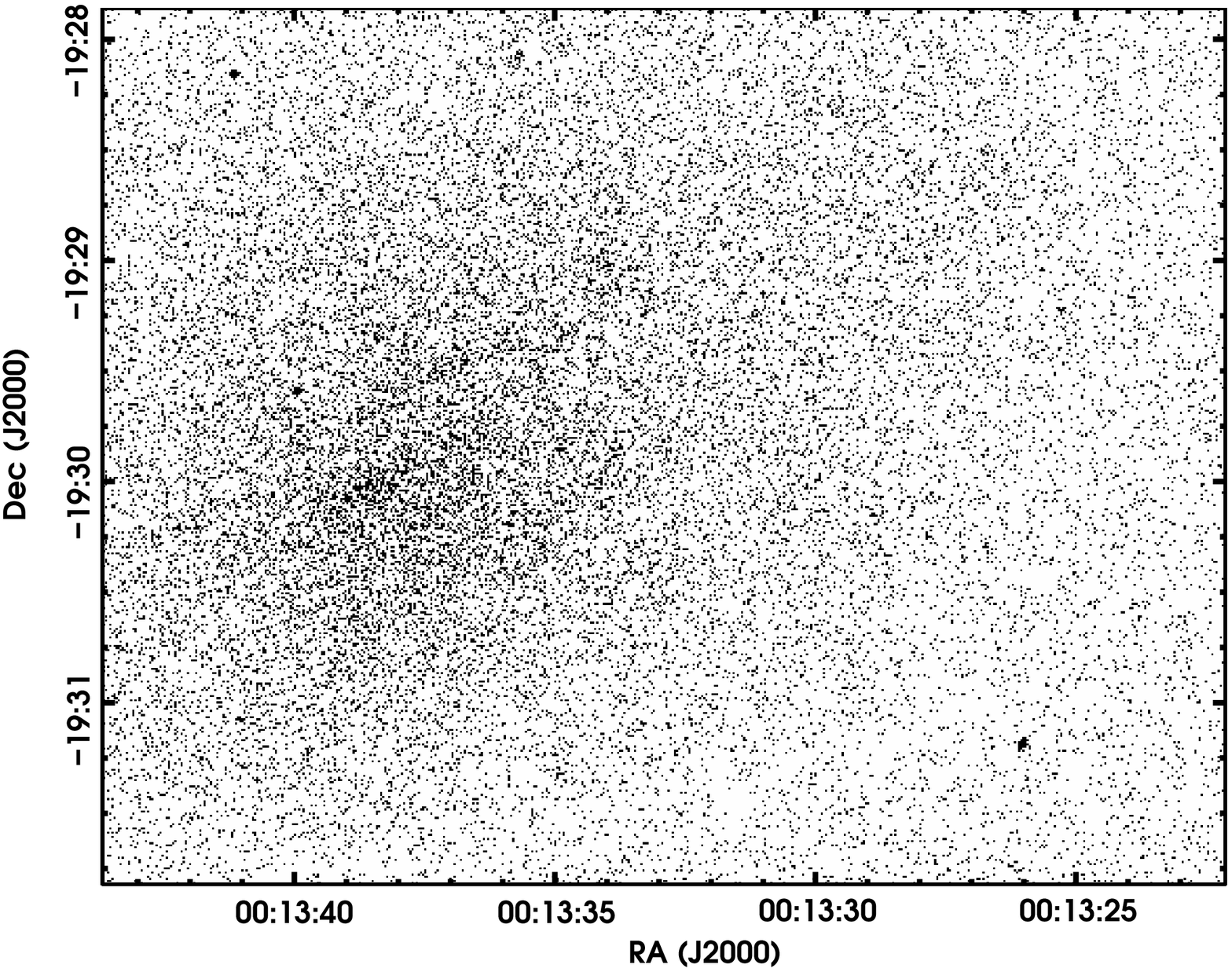,width=0.5\linewidth}\epsfig{file=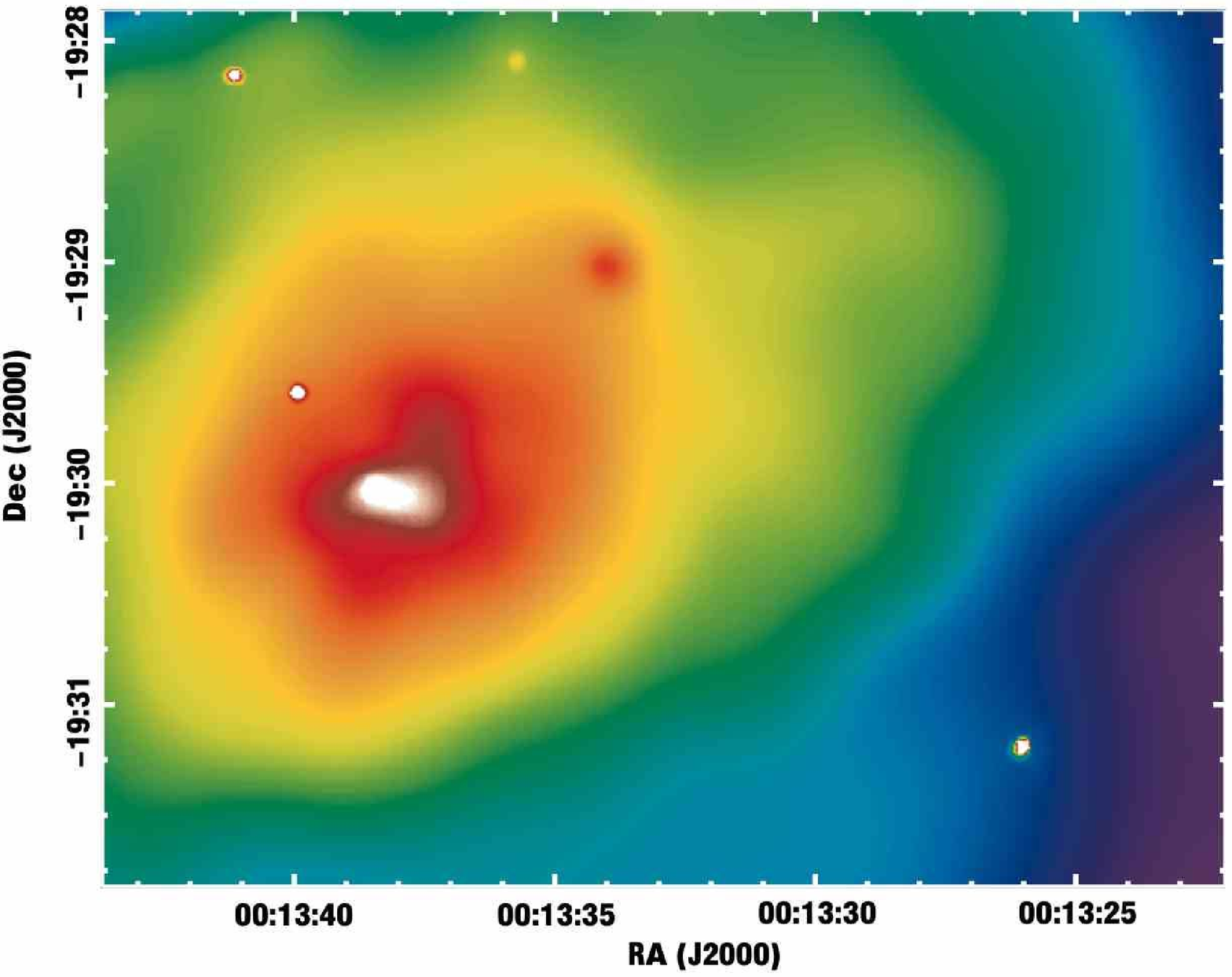,width=0.5\linewidth}}
\caption{{\em Left Panel:} Raw {\em Chandra} X-ray image (0.3--10~keV)
of the 5\arcmin $\times$ 4\arcmin\/ region of Abell 13 encompassing
the cluster center and radio relic positions.  The image is
uncorrected for exposure or background.  North is up, and east is
left.  The image pixels are 0\farcs492.  {\em Right Panel:} Adaptively
smoothed {\em Chandra} X-ray image of the same region, corrected for
background, exposure, and vignetting.  Notice the structure in the
X-ray core as well as the knot of emission to the northwest of the
core.  In addition, the X-ray emission is extended to the northwest
and west (towards the radio relic).  The number of point sources is
consistent with that expected from background sources.}
\label{fig:x}
\end{figure*}

In Figure~\ref{fig:x}, we show both the raw {\em Chandra} X-ray image
of the 0.3--10~keV energy range, and an adaptively smoothed image of
the X-ray data produced with the CIAO tool {\tt csmooth}.  The
smoothed image has a minimum signal-to-noise ratio of 3 per smoothing
beam, and has been corrected for exposure and background.  The X-ray
emission from Abell 13 shows complex structure in the core region.  A
knot of X-ray emission is also clearly detected 86\arcsec\/ (150~kpc)
to the northwest of the cluster's X-ray peak, along with a more
diffuse extension of the emission in the same direction.

We identify 11 point sources in our full band image (0.3--10.0~keV;
see Table~\ref{tab:srcprop}).  This is consistent with the expected
number of background sources from the {\em Chandra} Deep Fields
\citep{bah+01,grt+01}.  No point source was found at the center of the
X-ray emission.  Since the cluster emission will reduce our point
source detection sensitivity, we also searched for point sources in
just the hard band (4.0--10.0~keV) data.  In the hard X-ray band, the
diffuse emission from the cluster is weaker while the emission from
AGN is stronger than at softer bands.  Therefore, any AGNs should be
more easily detected.  We found no point sources within 0\farcm5 of
the cluster center, defined to be the position of galaxy H, the
brightest cluster galaxy.

\begin{figure*}
\centerline{\epsfig{file=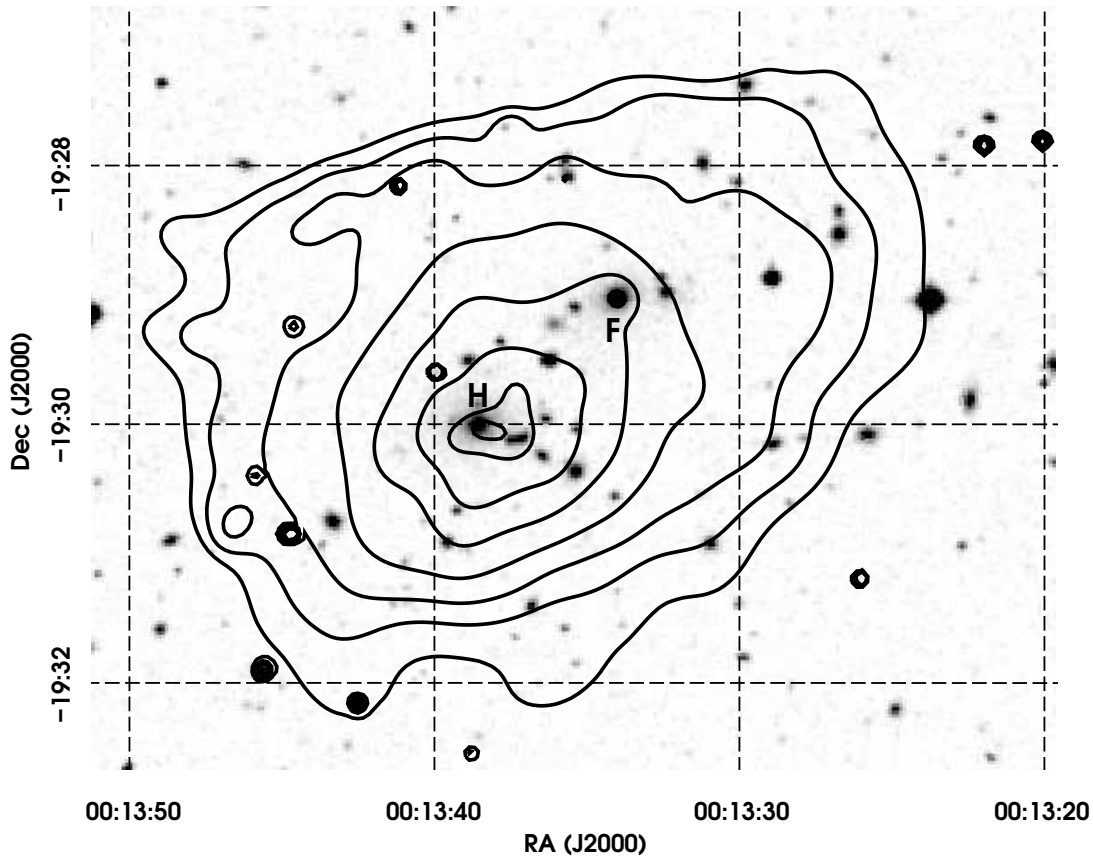,width=0.8\linewidth}}
\caption{X-ray brightness contours (0.3--10~keV band, logarithmically
spaced by a factor of $\sqrt{2}$) overlaid on the DSS optical image
scanned from a SERC-J survey plate taken in 1977 at the UK Schmidt
telescope, Australia.  The bright galaxies H and F from \citet{srm+01}
are labeled.  The X-ray emission is centered near galaxy H.  A knot of
X-ray emission to the northwest of the cluster X-ray peak is
consistent with the position of galaxy F.  The steepness of the X-ray
contours on the northern and eastern edges are artifacts of the CCD
edge.}
\label{fig:op}
\end{figure*}

In Figure~\ref{fig:op}, we show the optical Digital Sky Survey (DSS)
image\footnote{Available at http://archive.stsci.edu/dss/} of the
cluster with the X-ray contours overlaid.  The X-ray contours are
based on the image in Figure~\ref{fig:x} {\em right panel}.  We have
labeled the brightest galaxies following \citet{srm+01}.  The X-ray
emission is centered near the bright elliptical H (also 2MASX
J00133853$-$1930007), the brightest cluster galaxy.  The knot of
emission to the NW is coincident with the elliptical galaxy F (also
2MASX J00133401$-$1929017), the second brightest cluster galaxy.  The
steepness of the X-ray gradient to the north and east is an artifact
of the CCD edges.

\begin{figure*}
\centerline{\epsfig{file=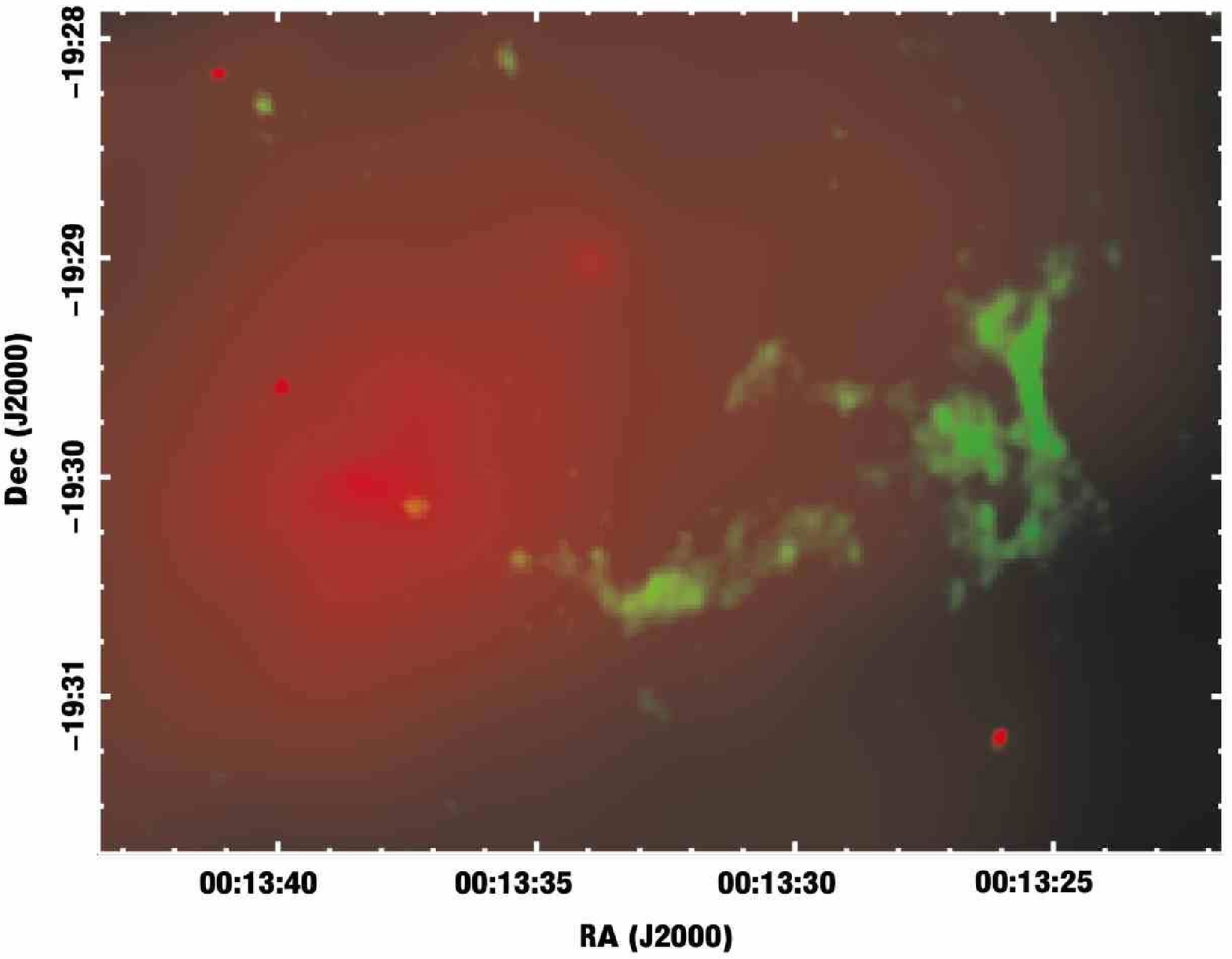,width=0.8\linewidth}}
\caption{Radio image at 1.4~GHz \citep{srm+01} overlaid in green the
adaptively smoothed X-ray image of Abell 13 in red.  North is up, and
east is left.  The region covered is 5\farcm1 $\times$ 3\farcm8.  The
radio image shows only emission detected at greater than 2.5$\sigma$
as determined by \citet{srm+01}.  The three compact red spots are
X-ray point sources unrelated to the cluster.}
\label{fig:xr}
\end{figure*}

We are also interested in comparing the X-ray data with the radio
emission from Abell 13.  Figure~\ref{fig:xr} shows the X-ray data in
red and the radio data in green.  We find no enhancement of the X-ray
emission near the radio relic, which is different from what was seen
for Abell 133 \citep{fsk+02}.  Given the low surface brightness of the
X-ray emission, it is difficult to determine if there is a deficit of
X-ray emission at the position of the radio relic, commonly referred
to as an X-ray cavity or radio bubble.  The absolute astrometric
accuracy of the radio and X-ray maps is roughly 0\farcs5 for each.
The radio point source near the cluster center is not associated with
galaxy H.  Instead, it seems coincident with a barely resolved pair of
galaxies in the DSS and Two Micron All Sky Survey (2MASS) images,
listed in NED\footnote{Available at http://newwww.ipac.caltech.edu/}
as 2MASX~J00133712$-$1930067.  No velocity information is available
for 2MASX~J00133712$-$1930067, making association with the cluster
uncertain.

\begin{figure*}
\centerline{\epsfig{file=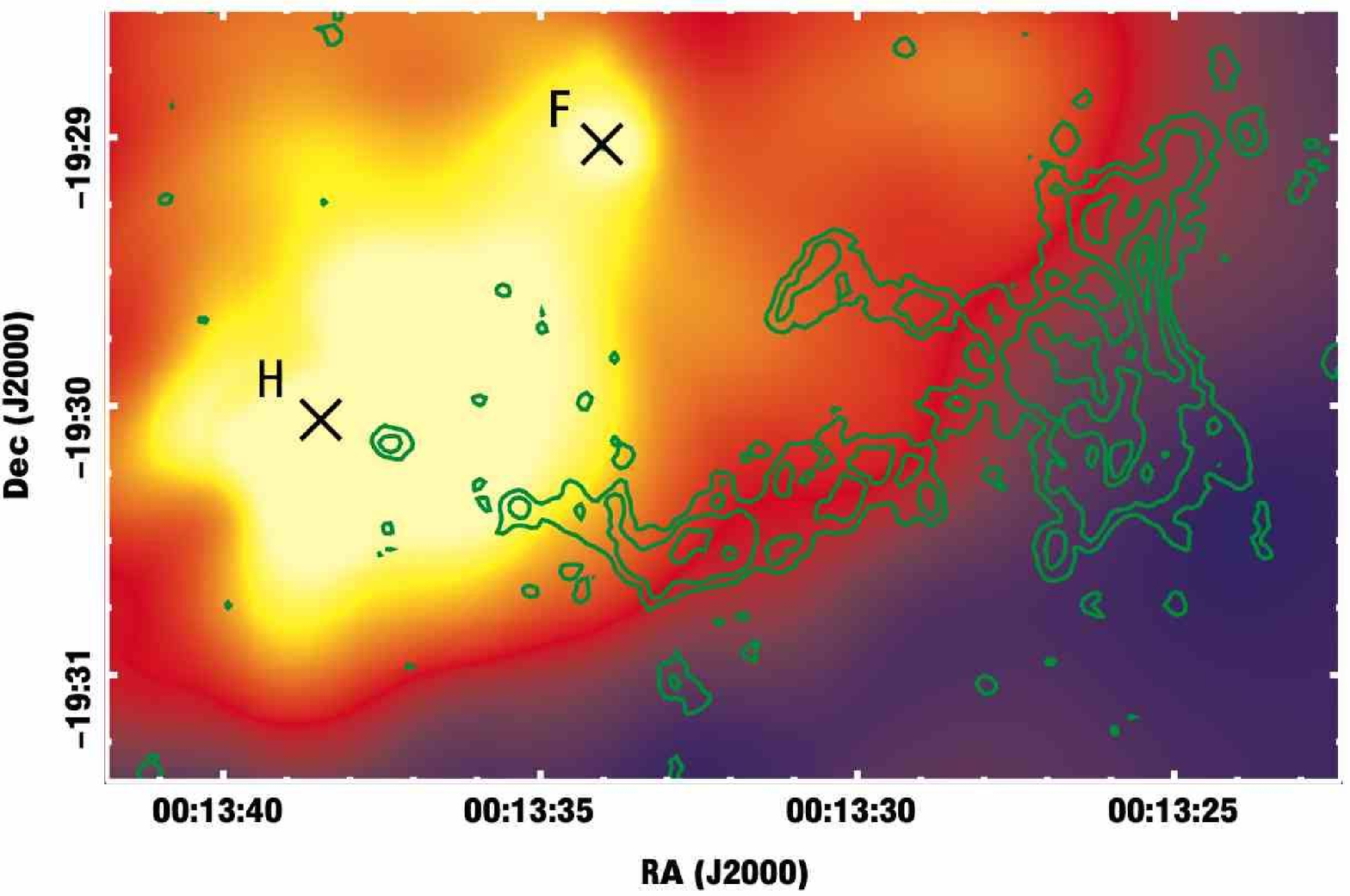,width=0.8\linewidth}}
\caption{Residual X-ray emission from the cluster (color scale from
blue$=$weak to white$=$strong) after partial subtraction of a smooth
elliptical isophotal model for the entire cluster.  The region shown
here encompasses the radio relic and has a size of 4\farcm6 $\times$
2\farcm8.  The color scale was chosen to emphasize the outer regions
of the cluster near the radio relic.  Overlayed are the radio contours
as given in \citet{srm+01} with contour levels of 43, 86, 173, 303 and
389 $\mu$Jy~beam$^{-1}$, which correspond to 2.5, 5, 10, 18 and 23
$\sigma$.  The two brightest galaxies H and F are indicated by the Xs.
No excess or clearly associated deficit in X-ray emission is found
coincident with the radio relic.  Along the eastern tail of the radio
emission there appears to be a sharp drop in the residual X-ray
emission.  However, this drop is not very strong in the total X-ray
surface brightness.  There is a strong gradient in the southern part
of the cluster, but at a larger radius and further south.}
\label{fig:comp}
\end{figure*}

To enhance the structures in the cluster core, we fitted the observed
cluster emission with a smooth elliptical isophotal model, using the
IRAF/STSDAS task {\tt ellipse}.  We allowed the ellipticity, position
angle, and intensities of each isophote to vary while keeping the
centroids fixed to the position of galaxy H.  The model was then
subtracted from the data after multiplication of the model counts by
0.5 to avoid oversubtraction.  In Figure~\ref{fig:comp}, we show the
subtracted image of the cluster in the region around the radio relic
with the radio contours overlaid.  No excess emission is associated
with the radio relic.  The cluster emission is faint near the radio
relic, and continues to weaken at projected cluster-centric distances
greater than that of the relic in the same direction.  

\begin{figure}
\centerline{\epsfig{file=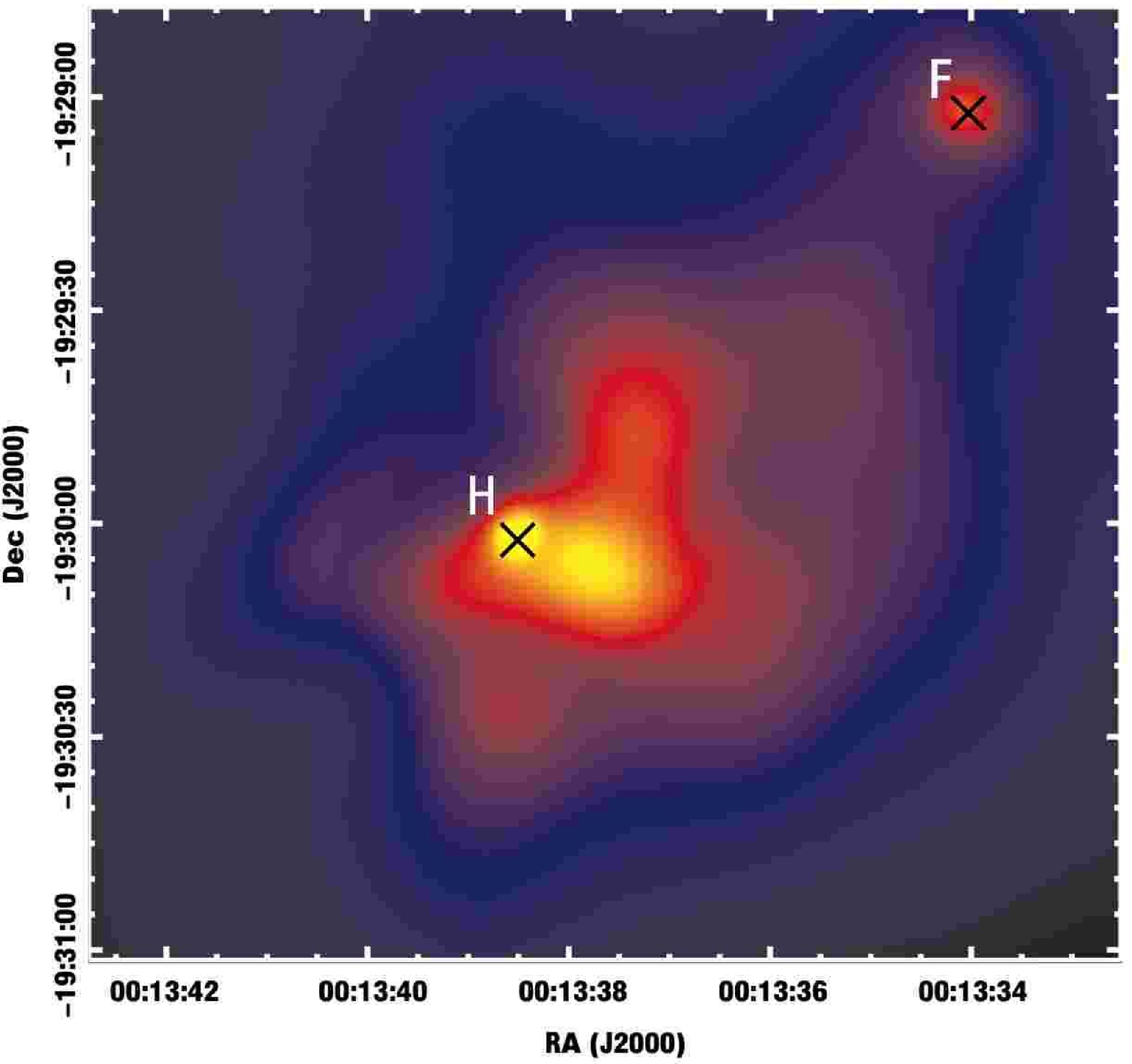,width=\linewidth}}
\caption{Residual emission from center of Abell 13 after partial
subtraction of a smooth elliptical isophotal model.  The figure spans
a region of 2\farcm4 $\times$ 2\farcm2.  The color scale was chosen to
emphasize the inner region of the cluster near galaxy H.  North is up
and east is left.  The bright region marked with an X near the center
coincides with galaxy H.  Notice the extension of emission to the west
(in the direction of the radio relic) and the excess emission
perpendicular to that direction.  The X at the top right corner marks
the position of galaxy F.  }
\label{fig:compcen}
\end{figure}

More interesting structure is seen in the cluster center.
Figure~\ref{fig:compcen} shows the same image as Figure~\ref{fig:comp}
with a different color scale to enhance the variations in the inner
region of the cluster.  The knot of emission in the upper right is the
second-brightest elliptical galaxy, F.  The bright yellow knot of
emission is coincident with galaxy H.  There is a more diffuse bright
extension of emission to the west of galaxy H.  In addition, there is
other structure in the cluster center perpendicular to the bright
extension.  Interestingly, the emission extension to the west of
galaxy H seems to point in the same direction as one might expect the
radio relic to extend if it were projected back to galaxy H.

\subsection{X-ray Emission from Galaxy F}
Enhanced X-ray emission has been seen in the large elliptical galaxies
in the Coma cluster \citep{vmf+01}.  We compared the enhanced emission
of galaxy F with the results from Coma.  First, we determined the
radial profile of the enhancement to determine its size and the
appropriate source region for galaxy F.  The emission was clearly more
extended than the {\em Chandra} point spread function.  We found that
the enhancement measured $\approx$1\farcs5--2\arcsec\/ in radius, or
roughly 3 kpc at the distance of Abell 13.  \citet{vmf+01} found that
the compact cores associated with the brightest galaxies in the Coma
cluster had a physical size of 3~kpc in radius.

We attempted to extract a spectrum of the enhanced emission associated
with galaxy F from a circular region with radius of 2\arcsec.  To
account for both the cosmic X-ray background and the cluster emission,
we used a local background extracted from an annulus centered on
galaxy F with radii of 3\arcsec\/ and 5\arcsec.  There are only 31
counts in a 2\arcsec\/ region around galaxy F, and we expect 17.5 of
these to be background.  This is too few counts for spectral fitting
or even to determine a useful color for the source.  Even a basic
color comparison cannot yield any constraint on the temperature of the
galaxy gas.

We used PIMMS to determine the flux from galaxy F.  We measured a
0.5--8.0~keV count rate of $2.1\times10^{-4}$ cts~s$^{-1}$ and assumed
a gas temperature of 1~keV for a Raymond-Smith plasma model.  The
largest elliptical galaxies have X-ray gas temperatures of
$\approx$1~keV \citep[e.g.,][]{opc03}.  From PIMMS we find a 0.5--2.0
keV unabsorbed flux of $8.2 \times 10^{-16}$~erg~cm$^{-2}$~s$^{-1}$,
or a luminosity of $2\times 10^{40}$ erg~s$^{-1}$.  The galaxies in
Coma had similar, although slightly larger (7.6--9.1$\times10^{40}$
erg~s$^{-1}$) luminosities.  We also calculated the mass of gas
necessary to produce the excess emission from galaxy F.  To do this,
we first calculated the model normalization of a 1~keV {\tt APEC}
plasma model that would produce the flux measured from galaxy F.  We
then calculated the average electron number density in the gas from
the model normalization and assuming a spherical gas distribution with
2\arcsec\/ radius.  Translating the electron number density into a gas
mass, we find that the required gas mass is $4.2 \times
10^{7}$~M$_{\odot}$.  This is comparable to the gas masses found in
the Coma galaxies and other galaxies in nearby clusters
\citep{vmf+01,sjf+07}.

Given the similarity in size and luminosity, we may expect that the
cause of the X-ray enhancement in galaxy F is the same as that found
for the Coma galaxies \citep{vmf+01}.  The gas in the Coma galaxies is
of low temperature (1--2~keV) and is likely the remains of the
galactic X-ray halos.  \citet{vmf+01} suggest that the cool gas is in
pressure equilibrium with the hotter, intercluster medium.

\section{Spectral Analysis}
Spectra were extracted from the level 2 event file, excluding point
sources within the regions of interest.  Background spectra were
extracted from the blank-sky background file.  For each spectrum,
weighted response files (RMFs and ARFs) were generated using the CIAO
tools {\tt mkacisrmf} and {\tt mkwarf}, respectively.  Finally,
spectra were grouped to obtain a minimum of 25 counts per bin.
Spectral fitting was primarily performed in XSPEC v11.0, except for
the temperature map spectral fits (see \S~\ref{sec:tmap} below).

\subsection{Total Spectrum}\label{sec:full}
\begin{figure}
\centerline{\epsfig{file=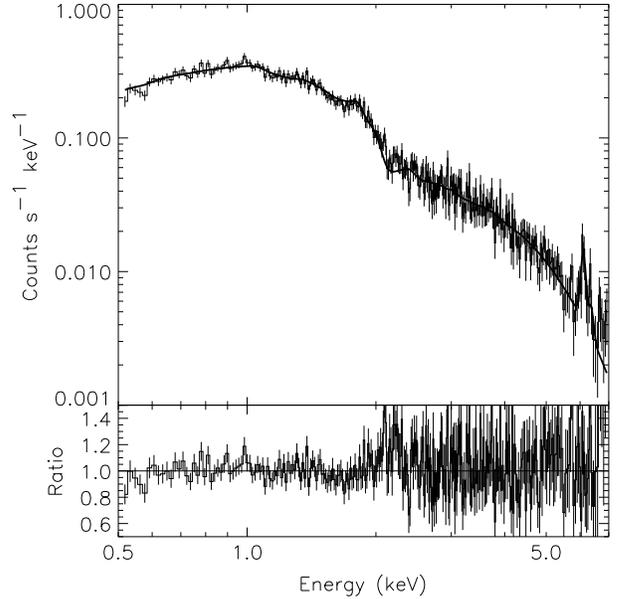,width=\linewidth}}
\caption{{\em Upper panel:} Count spectrum of the X-ray emission from
Abell 13 and the best fit model.  {\em Lower panel:} Ratio of the data
and best-fit model (data/model).  The spectrum was extracted from a
2\farcm5 $\times$ 1\farcm9 elliptical region encompassing the largest
elliptical isophote contained fully on the S3 CCD.  The best-fit model
has a temperature $kT = 6.0 \pm 0.3$~keV and a metal abundance of
$0.46\pm0.12$ solar.}
\label{fig:clusspec}
\end{figure}

We first determined the integrated spectrum of the cluster using an
elliptical region with semi-major and semi-minor axes of 2\farcm5 and
1\farcm9, respectively, and position angle of 121$\arcdeg$ measured
from north to east (see Figure~\ref{fig:specshape}).  This region was
the largest elliptical region whose shape agreed with the outer
cluster X-ray isophotes and fit entirely on the S3 CCD.  We fit the
data with an absorbed plasma model, using the {\tt APEC} model in
XSPEC.  We fixed the redshift of the cluster to the mean redshift of
Abell 13 found in the ENACS study \citep[$z=0.0943$;][]{mkd+96} and
the equivalent hydrogen column density to the average Galactic value
in the direction of Abell 13, $N_{\rm H} = 2.0 \times
10^{20}$~cm$^{-2}$ \citep{dl90}.  The best-fit model had a temperature
$kT = 6.0\pm0.3$ keV and a metal abundance of $0.46\pm0.12$ solar (see
Table~\ref{tab:spec} and Figure~\ref{fig:clusspec}).

We also performed a spectral study to determine if a cooling flow is
present in the center of Abell 13, i.e. the peak in the X-ray emission
which is also coincident with galaxy H.  We extracted spectra in
circular annuli out to 100\arcsec\/ from the cluster center.  Each
spectrum was required to have at least 1000 counts.  The width of the
annuli ranged from 15\arcsec\/ in the cluster center to 5\arcsec\/ at
the periphery of the cluster.  We fitted each spectrum with the same
{\tt APEC} model used to describe the total cluster emission, allowing
only the temperature and normalization to vary (the metallicity was
fixed at the best fit value found above).  No statistically
significant variations in temperature were found.  This result is
consistent with the temperature map determined in \S~\ref{sec:tmap}.

\subsection{Temperature Map}\label{sec:tmap}
\begin{figure*}
\centerline{\epsfig{file=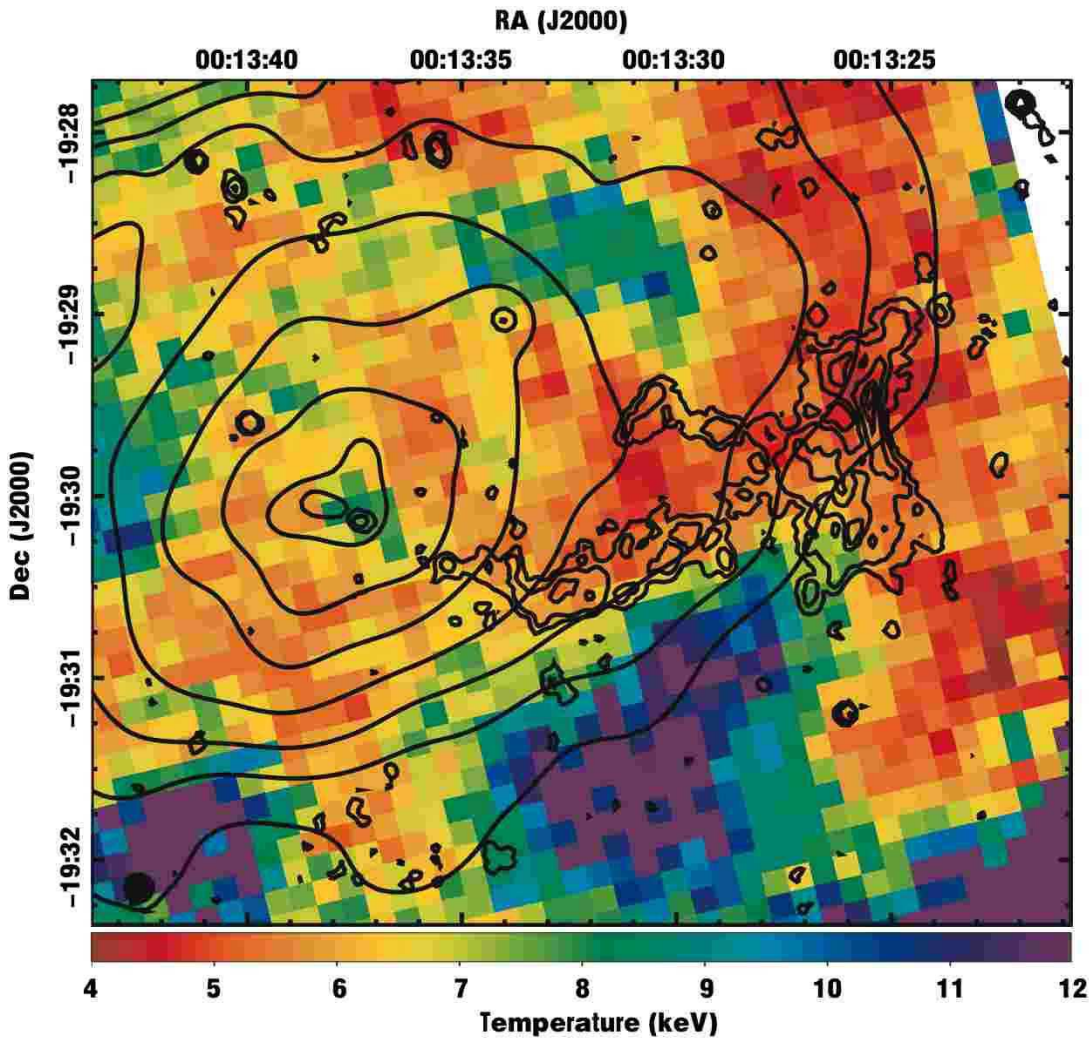,width=0.8\linewidth}}
\caption{Temperature map of Abell 13. Red signifies regions of lower
temperature, purple higher temperature.  Overlaid are the X-ray and
radio surface brightness contours.  No cooling flow is detected.  The
coolest gas is to the west of the peak in the X-ray emission, and
apparently associated with the radio relic.}
\label{fig:tmap}
\end{figure*}

In order to determine the spectral distribution in the cluster
emission, we created an adaptively binned temperature map of Abell 13,
shown in Figure~\ref{fig:tmap}.  The temperature map was produced
using ISIS \citep{hd00}.  A spectrum at each map pixel (1 map pixel
$=$ 16.3 ACIS pixels $=$ 8\farcs0) was extracted from the event data
(excluding point sources) in a square region centered on the pixel.
The region size was determined by requiring at least 800 net
(i.e. background subtracted) counts in the 0.4--7.0~keV energy range,
but with a maximum box size of 183$\times$183 ACIS pixels, or
90\arcsec$\times$90\arcsec.  The maximum box size was only reached in
the outskirts of the cluster.  The total counts in each region were
between 1.5 and 2 times the 800 count minimum to account for
background subtraction.  For each spectrum, a background spectrum was
extracted from the blank sky background event file in the same region.
RMF and ARF files were constructed for specific locations on the CCD,
with a spacing of 32 and 50 pixels, respectively.  This was done to
account for variations in the response across the CCD, without having
to create individual RMF and ARF files for each region which would be
computationally demanding.  For a given region, the RMF and ARF used
to fit the spectrum were chosen by selecting the response files with
positions closest to the count-weighted average position of the events
in each region.  The spectra were grouped to have 25 counts per bin
and were fitted in ISIS using the {\tt APEC} thermal model.  The
Galactic hydrogen column density, redshift, and abundance were fixed
to the values found for the full cluster spectrum (\S~\ref{sec:full}
above).  Only the temperature and normalization of the {\tt APEC}
model were allowed to vary.  In general, the errors on the temperature
map are about 20\% in the temperature.  Although, for pixels with
higher best-fit temperatures or those near the edge of the map, the
errors are larger, in the range of 50--70\%.  This is because of the
low high-energy sensitivity of {\em Chandra} and the lower statistics
of the outer regions.

The temperature map shown in Figure~\ref{fig:tmap} is colored such
that the coolest regions ($kT \approx 4$~keV) are red, and the hottest
regions ($kT \approx 12$~keV) purple.  The intensity contours for the
X-ray and radio emission from Abell 13 are overlaid.  We note that
small differences in temperature from pixel to pixel are not
statistically meaningful given the errors in the fits (an average
error of 1.2~keV at 6.0~keV).  Since the fitted region size is larger
than a single map pixel, there is overlap of the fitted regions.

Of particular interest are the lack of very cool gas in the cluster
center and the presence of cool gas associated with the radio relic
region.  The temperature map shows a large concentration of lower
temperature gas associated with the radio relic.  Below (\S
\ref{sec:rad}) we perform a detailed spectral study of this gas to
better constrain its properties.  There may also be some hotter gas
(purple/blue region) associated with the steep gradient in the X-ray
surface brightness contours on the southern edge of the cluster (see
e.g., the X-ray contours in Figure~\ref{fig:tmap}) but the temperature
errors are large in this region.  To try and confirm this result, we
extracted spectra on the northern and southern sides of the X-ray
gradient.  No significant difference in the best-fit temperatures was
found, although both regions had large errors.  Higher count spectra
are required to verify the increased gas temperature along the
cluster's southern edge.

\subsection{X-ray Spectral Study of the Radio Relic Region}\label{sec:rad}
\begin{figure}
\centerline{\epsfig{file=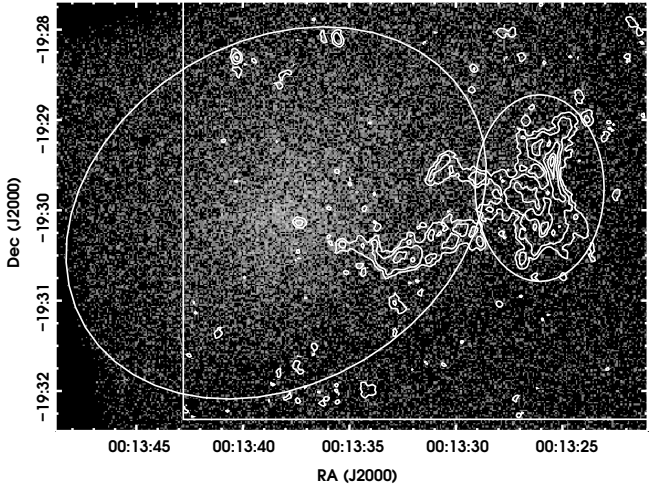,width=\linewidth}}
\caption{Raw {\em Chandra} X-ray image (0.3--10~keV) of Abell 13
encompassing the cluster center and radio relic positions.  Overlaid
are the radio contours of \citet{srm+01}.  The straight lines mark the
coverage of the VLA data compared to the {\em Chandra} data.  The
large ellipse marks the region of the data used to determine the
spectral model of the full cluster emission (\S~\ref{sec:full}).  The
smaller ellipse on the right marks the region of the data used to
measure the spectrum of the X-ray gas coincident with the radio relic
(\S~\ref{sec:rad}).  The region covered is 6\farcm6 $\times$
4\farcm7.}
\label{fig:specshape}
\end{figure}

From the temperature map, we see that the X-ray emission associated
with the radio relic is colder than the average cluster temperature.
This immediately rules out a shock origin as we would expect the
emission to be hotter in that case.  To determine the nature of this
cooler emission, we performed a more rigorous X-ray spectral study of
the radio relic region.  We fit the spectra from a
2\farcm0$\times$1\farcm4 (major/minor axis) elliptical region
surrounding the radio relic (Figure~\ref{fig:specshape}).  This region
was chosen to encompass the outer portion of the radio relic that is
associated with the cooler gas in the temperature map.  This region
has the advantage that it is a region of lower surface brightness for
the hot gas component.  If we add in the radio relic region to the
east (the tail leading back to galaxy H), the spectrum has a much
larger hot gas component and is less sensitive to a cool gas
component.

\begin{deluxetable}{lc}
\setlength{\tabcolsep}{0.2in}
\tablecaption{Best-Fit {\tt APEC} Model Parameters for
    Cluster\tablenotemark{a}}
\tablehead{\colhead{Parameter} & \colhead{Value}}
\startdata
$N_{\rm H}$ (cm$^{-2}$) & $2.0\times 10^{20}$\tablenotemark{b} \\
$kT$ (keV) & 6.0$\pm$0.3 \\
Abundance (solar) & 0.46$\pm$0.12 \\
Norm & (2.98$\pm$0.08)$\times 10^{-3}$ \\
$\chi^2$/dof & 316.45/297 \\
\enddata
\tablenotetext{a}{All errors quoted at the 90\%-confidence level.}
\tablenotetext{b}{Column density fixed at Galactic value.}
\label{tab:spec}
\end{deluxetable}

\begin{figure}
\centerline{\psfig{file=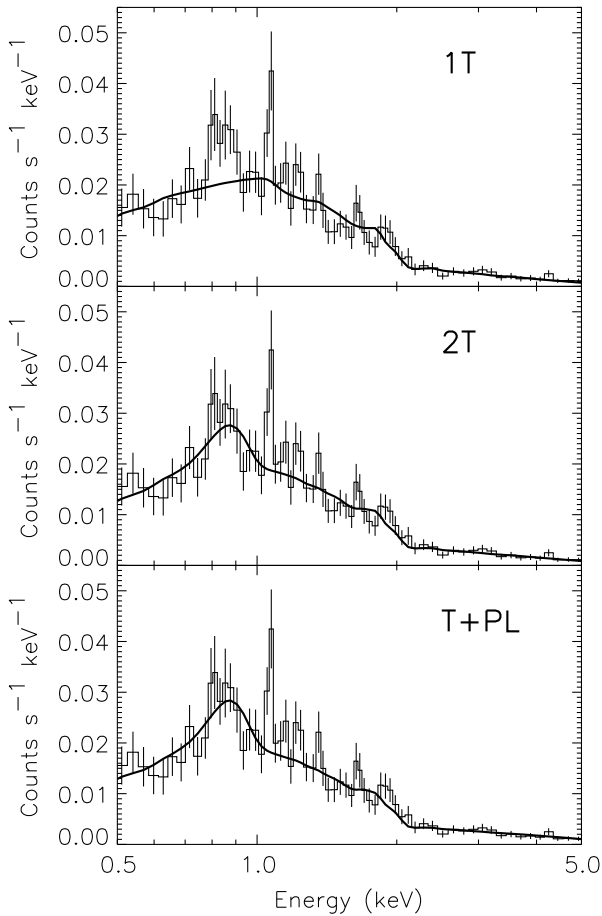,width=\linewidth}}
\caption{Count spectra of the X-ray emission associated with the radio
relic, fit with different models.  We have fit the spectrum with a
single temperature {\tt APEC} model ({\em top panel}), two {\tt APEC}
models ({\em middle panel}), and a powerlaw $+$ {\tt APEC} model ({\em
bottom panel}).  Notice the soft excess at 0.8~keV in the single
temperature model ({\em top}) but not the other spectra.  The peak at
$\sim$1.1~keV is too narrow to be astrophysical and is likely
strengthened by binning.  At other binnings, it is not as
significant.}
\label{fig:rrspec}
\end{figure}

The data were fit with different spectral models, including a single
temperature gas model, a two-temperature gas model, and a gas $+$
powerlaw emission model.  The results are given in
Table~\ref{tab:rrspec}.  In all fits we have frozen the metallicity of
the thermal components to that of the general cluster emission.  If
allowed to vary, the metallicity is unconstrained.  The single
temperature gas model, while a decent fit to the data, does not
reproduce the excess emission in the data near 0.8~keV (see
Figure~\ref{fig:rrspec}).  The two-temperature gas model is the best
fit to the data.  One of the components is consistent with the overall
cluster temperature, while the other component has a temperature of
$\approx$0.9~keV and accounts for the soft excess seen in the data.
The gas $+$ powerlaw model is a worse fit, with the powerlaw component
replacing the hotter gas component.  The two-temperature gas model is
the most appropriate, as it would include both a lower temperature gas
associated with the relic as well as hotter cluster emission seen in
projection.  

\section{Discussion}\label{sec:disc}
We have presented the {\em Chandra} data of the galaxy cluster Abell
13.  The X-ray emission is centered on galaxy H of \citet{srm+01}, the
brightest cluster galaxy.  The spatial structure near the core is
complicated, with a bright tail of excess emission to the west of
galaxy H, and a fainter excess to the north.  No point source is
associated with galaxy H and no cooling flow is found.  We also found
excess X-ray emission associated with the second-brightest elliptical
galaxy, F, in the cluster.  The mass of gas associated with galaxy F
is consistent with other bright elliptical galaxies in cluster
environments \citep{vmf+01,sjf+07}.  In general, there is an extension
of the X-ray emission to the northwest with respect to the X-ray peak
coincident with galaxy H.  Finally, no excess or significant deficit
in emission is found coincident with the unusual radio relic in Abell
13, although the spectral properties of the gas in this region are
different from those in other areas of the cluster.  We found a soft
excess in the spectrum of gas coincident with the radio relic, which
can be well fit by a $\approx$0.9~keV thermal model, suggesting the
presence of cooler gas there.

The redshift distribution of the galaxies in Abell 13 can be described
by a bimodal distribution \citep{fgg+96}.  The subclumps are not
distinguishable in their location on the sky.  Interestingly, the two
brightest galaxies, H and F, are associated with different peaks in
the redshift distribution and rather than being associated with the
peak velocity, either of the cluster as a whole or of either subclump,
the velocities of galaxies H and F are on the very edges of the
velocity distribution.  Galaxy F is on the lower edge for the
low-velocity subcluster and galaxy H on the higher edge of the
high-velocity subcluster.  This, combined with the complicated
structure of the X-ray emission, points to an ongoing merger within
the cluster and that galaxies H and F have a significant velocity
relative to the general cluster distribution.  The spatial
distribution of the galaxies in Abell 13 shows a northwest-southeast
extension, paralleling the X-ray distribution, but extending to larger
scales.

In the following discussion, we assume that the X-ray gas associated
with the radio relic has two components, a hot (6~keV) gas component
due to the general cluster emission, and a cooler (0.9~keV) gas
component local to the relic.  This model is based on our best-fit
spectrum found in \S~\ref{sec:rad}.  We assume that the two components
are in pressure equilibrium.

From \citet{srm+01}, the minimum pressure of the radio relic in Abell
13 is $1.4 \times 10^{-12}$ dyne cm$^{-2}$.  Their estimate of the
thermal pressure of the hot X-ray gas was very rough and based solely
on scaling the properties of Abell 85 since no imaging observation of
Abell 13 was available.  Our {\em Chandra} data allow for a more
accurate estimate of the thermal pressure at the location of the
relic.  However, due to an unfortunate placement of the cluster on the
CCD, the only regions which extended out to a distance larger than the
relic were in the direction of the relic and to the south.  Since the
southern edge of the cluster shows a steep gradient in the X-ray
emission, we did not consider this region in determining the surface
brightness distribution.  Instead, we determined the surface
brightness of the cluster in a conical region extending from the
cluster center out to 4\farcm1 and encompassing position angles of
240--312$\arcdeg$ measured from the north to the east.  This region
includes the radio relic and associated cool gas which will affect our
results; however, while not a perfect measure of the properties of the
X-ray gas, it represents a significant improvement on the estimates of
\citet{srm+01}.  A larger field of view observation would be better
suited to address the surface brightness and electron density
distributions of the full cluster emission.

\begin{deluxetable*}{lccccc}
\tabletypesize{\footnotesize}
\tablewidth{0pt} 
\setlength{\tabcolsep}{0.1in}
\tablecaption{Best-Fit X-ray Spectral Model Parameters for Radio 
   Relic\tablenotemark{a}}
\tablehead{\colhead{Model} & \colhead{$kT_1$ (keV)} &
   \colhead{Norm$_1$} & \colhead{$kT_2$ or $\Gamma$} &
   \colhead{Norm$_2$} & \colhead{$\chi^2$/dof}}
\startdata
{\tt APEC} & 5.9$^{+1.9}_{-1.1}$ & (1.77$\pm$0.08)$\times 10^{-4}$ & 
     \nodata & \nodata & 89.4/69 \\
{\tt APEC} $+$ {\tt APEC} & 0.86$^{+0.16}_{-0.09}$ & 
     (1.5$\pm$0.8)$\times 10^{-5}$ & 
     10$^{+12}_{-4}$ & (1.60$\pm$0.12)$\times 10^{-4}$ & 74.3/67 \\
{\tt APEC} $+$ PL & 0.89$^{+0.16}_{-0.09}$ & (1.8$\pm$0.8)$\times 10^{-5}$ & 
     1.40$\pm$0.08 & (3.2$\pm$0.3)$\times 10^{-5}$ & 81.4/67 \\
\enddata
\tablenotetext{a}{All errors quoted at the 90\%-confidence level.  All
models were fit including Galactic absorption of $2.0\times
10^{20}$~cm$^{-2}$. Abundances for gas emission models were fixed to
the best-fit value from the cluster spectrum, 0.46 of solar.  The
model parameters Norm$_1$ and Norm$_2$ are the values of the first and
second model component normalizations.  $kT_2$ or $\Gamma$ is the value
of the second model component temperature if {\tt APEC} or photon index
if powerlaw.}
\label{tab:rrspec}
\end{deluxetable*}

Using the surface brightness distribution, we estimate the electron
density distribution of the hot component following the spherical
deprojection method of \citet{kcc83}.  We assumed a single temperature
of the hot gas of 6.0~keV and used the {\tt APEC} thermal model to
convert from counts to electron density.  From this, we find an
electron density at the relic of $1\times10^{-3}$~cm$^{-3}$.  Given
this and the average cluster temperature (6.0~keV) determined in
\S~\ref{sec:full}, we find that the thermal pressure of the
surrounding hot gas at the location of the radio relic is $1.6 \times
10^{-11}$ dyne cm$^{-2}$.  This value is larger than the estimate for
the pressure of the radio plasma, as has been seen in other X-ray
cluster radio bubbles \citep{dft05}.  We note that if the relic were
at a larger radius than the projected radius, the electron density and
thermal pressure would be lower, bringing it closer to the radio
plasma pressure estimate.  In addition, in the case of Abell 13, if
the values assumed by \citet{srm+01} for (1) the ratio of masses in
heavy particles and electrons; (2) the plasma filling factor; (3) the
low frequency spectral index, and (4) the angle between the line of
sight and field direction are given more realistic values then we may
increase the radio plasma pressure to that of the confining X-ray gas.

\subsection{Origin of the Radio Relic}
It is unlikely that the radio relic in Abell 13 is a peripheral
cluster relic associated with a cluster merger shock.  First, there is
no evidence in the {\em Chandra} image or spectra for any shock
feature near the relic.  Second, the radio image shows a long filament
which extends almost all the way back to the brightest elliptical
galaxy H (Figure~\ref{fig:xr}).  Thus, we assume that the radio
emitting plasma originated in an AGN at the center of galaxy H.  Based
on the classification system of \citet{kbc+04}, which looks at the
physical properties and proposed origins of extended cluster radio
sources, we would classify Abell 13 as an AGN relic system.

We propose two explanations for the multi-wavelength properties of
Abell 13.  The first is the buoyant bubble scenario suggested by
\citet{fsk+02} for Abell 133.  The radio source may have uplifted the
cool gas from the core of Abell 13.  In this scenario, galaxy H would
have been an active radio emitter in the past.  At some later time,
the radio plasma would have risen by buoyancy, and entrained the cool
gas in the center, uplifting it to a larger distance from the cluster
center.  This might also remove the source of fuel for the central
AGN.  This would explain the lack of cool gas in the cluster center
and the fact that currently, galaxy H is neither a radio
\citep[$<0.1$~mJy at 1.4~GHz;][]{srm+01} nor a hard X-ray source.  To
estimate the time scale for the cool gas and radio source to rise by
buoyancy from the position of galaxy H, we assume that the cool gas
and radio source move at a velocity roughly 1/3 the sound speed,
$c_s$, of the hot gas component.  For a 6.0~keV gas, $c_s$ is
980~km~s$^{-1}$.  The projected separation between the relic and
galaxy H is 2\farcm2--3\farcm1.  This yields a buoyancy time scale of
7--$10 \times 10^{8}$~yr, roughly consistent with the relic age
determined through spectral fitting \citep{srm+01}.  To estimate the
mass of cool gas uplifted, we take the volume to be that of a prolate
spheroid with major and minor axes equal to that of the region used to
extract the spectrum of the radio relic.  From the two-temperature
spectral model, we find that the enclosed mass of cool gas is $\sim 2
\times 10^{10}$~M$_{\odot}$.  This is a factor of 10 greater than the
mass of gas contained in the tongue of X-ray gas found in Abell 133
\citep{fsk+02}.  However, we note that the electron density for the
cool gas is $\sim 5 \times 10^{-4}$~cm$^{-3}$, lower than the hotter
cluster gas (particularly the gas towards the center of the cluster)
making buoyant uplift a possibility.  A cool gas mass of $\sim 2
\times 10^{10}$~M$_{\odot}$ implies that previously Abell 13 would
have had a mass deposition rate of $\sim 2$~M$_{\odot}$~yr$^{-1}$
assuming a cooling time of $10^{10}$~yr.  This is on the low end of
the distribution of known cooling flow mass deposition rates
\citep[see e.g.,][]{wjf97}.  This is only an estimate of the mass of
cool gas to determine if the value is realistic.  The temperature map
shows additional cool regions to the north and east of the radio relic
which if taken into account would increase the mass of cool gas.
However, the radio relic is filamentary in structure and it is
possible that the cool gas also resides in filaments rather than a
single big blob, which would decrease the gas mass estimate.  Since it
is difficult to determine the extent of either of these situations, we
merely suggest that the estimated amount of cool gas in Abell 13 seems
reasonable for a galaxy cluster.

The lower temperature of the gas near the radio source might be due
both to the lower entropy of the gas in the cool core of the cluster,
plus the effects of adiabatic expansion as this gas was moved out to a
lower-pressure region.  From our above estimate of the electron
density distribution of the cluster gas, we find that the electron
density at the cluster center is $\approx$6 times that at the radio
relic.  The pressure difference between the two locations should also
be 6 times assuming a constant hot gas temperature.  We assume that
the cool gas component is always in pressure equilibrium with the hot
gas component that dominates the cluster emission.  As a result of
adiabatic expansion, the temperature of the cool gas in its current
location, $T_{2}$ is related to its initial temperature, $T_{1}$, by
\begin{equation}
T_2 = T_1 \left( \frac{P_2}{P_1} \right)^{(\gamma - 1)/\gamma}
= T_1 \left( \frac{P_2}{P_1} \right)^{2/5}
\, ,
\end{equation}
where $P_1$ is the pressure in the cluster core and $P_2$ is the
pressure at the location of the radio relic, and the adiabatic index
is $\gamma = 5/3$.  Observationally, we found $P_1/P_2 \approx 6$,
giving $T_2 \approx 0.5 \, T_1$ for adiabatic expansion.  This
difference alone cannot account for the cooling of the gas from the
average cluster temperature of $\approx$6~keV to its current
temperature of $\approx0.9$~keV.  Therefore, the cool gas associated
with the radio relic must have been cooler than the average cluster
temperature to begin with.  We can estimate the initial cool gas
temperature assuming that adiabatic expansion was the only heating or
cooling mechanism as the cool gas was uplifted.  This requires that
the initial temperature of the cool gas was about $T_1 \approx 2$~keV
when this gas was located at the cluster core.  This is a reasonable
temperature for a cooling core.  If the relic were at a larger radius
than the projected radius, then we would expect the electron density
to be lower and therefore the adiabatic cooling to be stronger.  This
would make the initial temperature of the cool gas higher, possibly
more in line with the average cluster temperature.

Our second proposed explanation is that a dynamical event is the
origin of the unusual radio relic and disturbed X-ray morphology.
Both the X-ray and optical properties (radial velocity distribution)
suggest that Abell 13 is undergoing a merger event.  The X-ray
extension from galaxy H (to the west) points back to the radio relic.
This may imply that galaxy H is moving to the east and that the radio
relic is at the former position of the galaxy.  (We note that the
extension may also favor the buoyant uplift scenario.)  We propose
that before the merger event, galaxy H had a cool core and showed AGN
activity.  As the merger progressed, the cold core and radio lobes
were displaced (or stripped) from the galaxy as a result of ram
pressure from the gas in the other subcluster, which did not affect
the stars and dark matter in galaxy H.  \citet{srm+01} found that the
travel time from galaxy H was comparable to the estimated relic age.
The long filament of radio emission leading back almost to galaxy H
(Figure~\ref{fig:xr}) suggests that the AGN remained active during the
initial phases of this stripping process.  The concentration of X-ray
emission around galaxy H (Figure~\ref{fig:compcen}) suggests that the
stripping process is not complete; there is still gas around galaxy H,
but it is relatively hot, perhaps due to heating processes associated
with the merger.  The {\em Chandra} data do not conclusively determine
which of these two models is correct, or strongly reject either.

\acknowledgements{We would like to thank Gregory Sivakoff and Marios
Chatzikos for useful discussions.  We would also like to thank the
anonymous referee whose comments helped to improve the paper.  Support
for this work was provided by the National Aeronautics and Space
Administration primarily through {\it Chandra} award GO4-5133X, but
also through GO4-5137X and GO5-6126X, issued by the Chandra X-ray
Observatory, which is operated by the Smithsonian Astrophysical
Observatory for and on behalf of NASA under contract NAS8-39073.  Some
support also came from NASA XMM-Newton award NNG04GO34G.  Basic
research in radio astronomy at NRL is supported by 6.1 Base funding.
HA acknowledges support from CONACyT grant 40094-F.  Y.~F.\ is
supported in part by a Grant-in-Aid from the Ministry of Education,
Culture, Sports, Science, and Technology of Japan (17740162).
L.~R. acknowledges support from the National Science Foundation under
grants AST-0307600 and AST-0607674 to the University of Minnesota.}

\end{document}